\documentclass[floatfix,intlimits,pra,showkeys,showpacs,twocolumn]{revtex4-1}

\usepackage{amsfonts,amsmath,amssymb}
\usepackage{dcolumn}
\usepackage[dvips]{graphicx}
\usepackage{hyperref}

\begin{document}

\title{Multistage Zeeman deceleration of atomic and molecular oxygen}

\author{Theo Cremers}
\author{Simon Chefdeville}
\author{Vikram Plomp}
\author{Niek Janssen}
\author{Edwin Sweers}
\author{Sebastiaan Y.T. van de Meerakker}
\email{basvdm@science.ru.nl}

\affiliation{Radboud University, Institute for Molecules and Materials, Heijendaalseweg 135, 6525 AJ Nijmegen, the Netherlands}

\date{\today}

\begin{abstract}
Multistage Zeeman deceleration is a technique used to reduce the velocity of neutral molecules with a magnetic dipole moment. Here we present a Zeeman decelerator that consists of 100 solenoids and 100 magnetic hexapoles, that is based on a short prototype design presented recently [Phys. Rev. A  95, 043415 (2017)]. The decelerator features a modular design with excellent thermal and vacuum properties, and is robustly operated at a 10 Hz repetition rate.  We use this decelerator to demonstrate for the first time the state-selective deceleration of atomic oxygen to final mean velocities in the 500 - 125 m/s range. We characterize our decelerator further with molecular oxygen, which despite its heavier mass is velocity tuned in the 350 - 150 m/s range. This corresponds to a maximum kinetic energy reduction of 95\% and 80\% for atomic and molecular oxygen, respectively. The long multistage Zeeman decelerator presented here demonstrates that the concept of using alternating hexapoles and solenoids is truly phase stable. This Zeeman decelerator is ideally suited for applications in crossed beam scattering experiments; the state-selected and velocity controlled samples of O atoms and O$_2$ molecules are particularly relevant for studies of inelastic and reactive processes.
\end{abstract}

\pacs{37.10.Mn, 37.20.+j}
\maketitle

\section{Introduction}\label{sec:intro}
Controlled scattering experiments have proven indispensable in testing atomic and molecular interaction models \cite{Vogels:PRL113:263202}. To control the initial conditions, these types of experiments can benefit from velocity-controlled and state-pure molecular beams. The particles of interest often carry no charge, making their preparation for scattering experiments challenging. However, when the particles of interest have an electric or magnetic dipole, the dipole interaction with electric or magnetic fields can be used to manipulate the velocity and/or state distributions of these neutral particles. With the Stark decelerator technique, for instance, high-density beams of neutral polar molecules with narrow velocity spread are created and used in high-resolution scattering experiments \cite{Onvlee:PCCP16:15768, Onvlee:nchem9:226}. Inhomogeneous electric fields have also been used to contain molecules in a molecular synchrotron, providing high density packets of ND$_3$ that collide with argon \cite{Poel:PRL120:033402}. \\

While beams of paramagnetic particles have been manipulated with magnetic fields since the experiment of Stern and Gerlach \cite{Gerlach:ZP8:110}, the use of time-varying magnetic fields to slow down molecules in so-called Zeeman decelerators is relatively new. This technique was pioneered by the Merkt group at the ETH Z\"urich \cite{Vanhaecke:PRA75:031402} and the Raizen group at the University of Texas \cite{Narevicius:NJP9:358} in 2007. These decelerators consist of multiple single-pulsed solenoids used to slow down and focus paramagnetic particles. They provide excellent control over the state purity and longitudinal velocity of a beam, and have so far been used in novel spectroscopic \cite{Jansen:PRL115:133202, Jansen:JMolSpec322:9} and trapping experiments \cite{Hogan:PRL101:143001, Wiederkehr:PRA81:021402}. More recently, so-called traveling wave Zeeman decelerators have been constructed \cite{Lavert-Ofir:PCCP13:18948,Trimeche:EPJD65:263}, in which a moving and velocity tunable magnetic trap is created by overlapping quadrupole magnetic fields. These decelerators have successfully been used to confine high density samples of atoms and molecules in magnetic traps \cite{Liu:PRL118:093201, Liu:PRA91:021403, Akerman:PRL119.073204}.\\

Here, we present a 3-meter long multistage Zeeman decelerator that is developed and optimized in  particular for applications in crossed beam scattering experiments. The decelerator is an improved and extended version of a short prototype that we presented recently to decelerate metastable helium \cite{Cremers:PRA95:043415}. The unique feature of this decelerator concept is the use of an alternating array of solenoids and hexapoles, that allow for independent control over the longitudinal and transverse velocity of the beam. The new decelerator features a modular design, contains 100 solenoids and 100 magnetic hexapoles, and has excellent cooling and vacuum properties that allows for robust operation at a 10 Hz repetition rate.
We use this new decelerator to produce state-selected beams of O ($^3$P$_2$) atoms,  a species that had not been used in Zeeman deceleration before, which we tune to a final mean velocity in the 500 - 125 m/s range. We further characterize the decelerator using beams of O$_2$ (X$^3\Sigma_g^-$) molecules, which we tune to velocities in the 350-150 m/s range.
For both O and O$_2$, excellent agreement is obtained with the arrival time distributions predicted by numerical trajectory simulations of the beam manipulation process, indicating that our decelerator concept of using and alternating sequence of hexapoles and solenoids is truly phase stable. This allows for the production of intense, state-selected and velocity tunable packets of paramagnetic species, that are ideally suited for controlled elastic, inelastic or reactive scattering experiments.

\section{Experimental setup}
\subsection{The molecular beam}
\begin{figure*}[!htb]
    \centering
    \resizebox{0.9\linewidth}{!}
    {\includegraphics{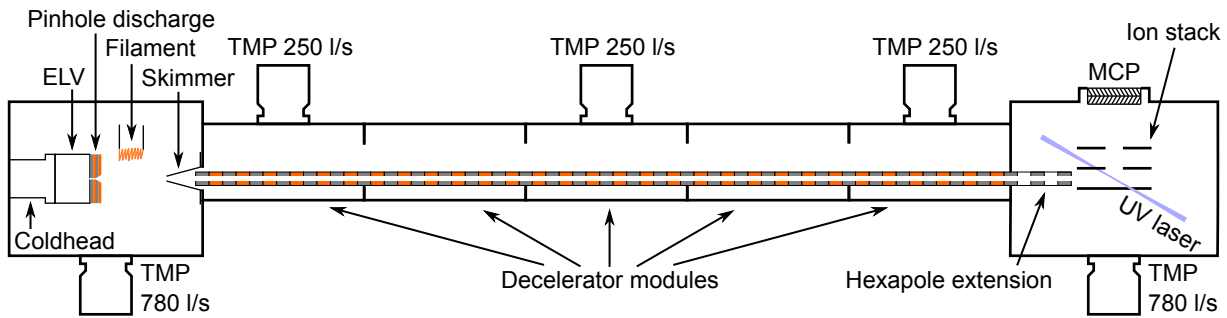}}
    \caption{Schematic representation of the experimental setup. The main components that are represented: the pulsed valve with optional discharge, the decelerator modules with turbomolecular pump (TMP) and the REMPI detection setup. A second source chamber (not shown here), is connected directly to the detection chamber, without the five decelerator modules in between. The sequence of alternating hexapoles and solenoids is represented by the (grey-orange) dashed lines inside the decelerator modules.}
    \label{fig:SchemSetup}
\end{figure*}
The setup is shown schematically in FIG \ref{fig:SchemSetup}. The molecular beam is created with a modified Even-Lavie valve (ELV) \cite{Even:EPJ2:17}. The default nozzle was replaced by a pinhole discharge that was used in our previous experiment with metastable He \cite{Cremers:PRA95:043415}. The reservoir behind the ELV is filled with a gas mixture of 10\% O$_2$ in Ar at a pressure of approximately 6 bar and the valve body is cooled to 200 K using a commercially available cold head (Oerlokon Leybold). At this temperature, the argon seeded molecular beam is created with a mean forward velocity of about 510 m/s. With an electrical discharge, the beam of molecular oxygen can be partially dissociated to atomic oxygen. In the $^3$P$_2$ ground state, oxygen atoms have a magnetic moment of three Bohr magnetons, and a mass of 16 amu. This magnetic moment to mass ratio is higher than that of ground state oxygen molecules (X$^3\Sigma_g^-$, N = 1), with a maximum magnetic moment of two Bohr magnetons and a mass of 32 amu. \\
The pinhole discharge is operated with a 50 $\mu$s pulse of -750 V on the front plate, thus creating an electrical discharge through the expanding gas, resulting in the dissociation of O$_2$ into its atomic components. The discharge was reliably operated by utilizing a 0.3 mm diameter tungsten filament running a 2.5 A current, positioned next to the ELV nozzle. We observe no significant velocity change in the molecular beam from the discharged current. After the supersonic expansion from the ELV and 100 mm of free flight, the beam is collimated by a skimmer (Beam Dynamics, model 50.8) with a 3 mm aperture. After the skimmer, the beam enters a series of modular vacuum chambers that contain sections of the multistage Zeeman decelerator.\\
Upon entering the decelerator chamber, the molecular beam immediately encounters the first magnetic hexapole in a series of alternating hexapoles and solenoids. The solenoids and hexapoles have an inner diameter of 3 mm. A secondary source chamber is connected directly to the detection chamber, separated by another (Beam Dynamics) skimmer. Beams created in this chamber are used to measure the state distributions of atomic and molecular oxygen without the influence of the multistage Zeeman decelerator. For these measurements, the ELV was moved from the primary to the secondary source chamber, to generate the same molecular beam.

\subsection{The multistage Zeeman decelerator}
\begin{figure}[!htb]
    \centering
    \resizebox{1.0\linewidth}{!}
    {\includegraphics{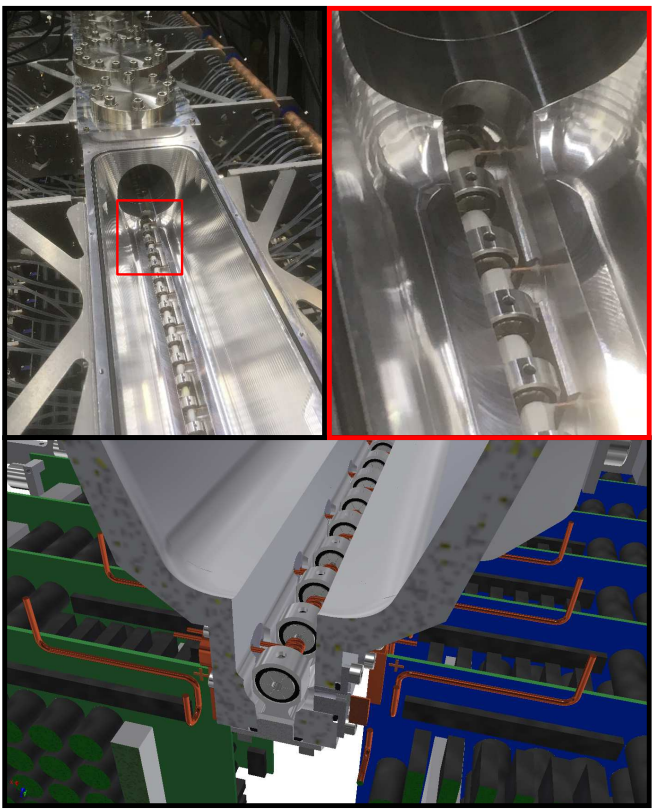}}
    \caption{Two photographs and a schematic cross-section of the multistage Zeeman decelerator used in this experiment. In the photographs, the lid is removed from the first vacuum chamber. The zoomed-in image shows the connection between two modules, which enables an uninterrupted sequence of solenoids and hexapoles. Solenoids are encased in white epoxy resin (Torr Seal) to resist flexing forces in the solenoids due to self-inductance. The cross-section shows the close proximity of the solenoids and the circuit boards that drive the solenoid currents.}
    \label{fig:DecZoom}
\end{figure}
A new multistage Zeeman decelerator was constructed based on the same technology featured in the prototype design \cite{Cremers:PRA95:043415}. This prototype design had the following conceptual differences compared to other multistage Zeeman decelerators:\\
First, the solenoids are made from copper capillary, and consist of only four windings. This solenoid has a much lower resistance compared to the solenoids used in other Zeeman decelerator designs that consists of multiple windings of thin copper wire. With this low resistance, the switching of high currents is achieved using low operation voltages, which allows us to substitute expensive IGBT-based high-voltage components with their cheap FET-based low-voltage counterparts. In the solenoid an on-axis magnetic field of approximately 2 T is produced by pulsing a 4.5 kA current for around 30 $\mu$s at repetition rates of 10 Hz, using an operation voltage of 24 V. The solenoid temperature is kept below 30 C$^{\circ}$ by passing cooling liquid through the solenoid capillary.\\
The second difference is the addition of magnetic hexapoles between each of the solenoid elements. Each hexapole consists of 6 permanent magnets in a ring shaped configuration with alternating magnetization pointing radially inward and outward. This configuration produces a magnetic field that is zero on the beam axis, and about 200 mT at the maximum radial position of 1.5 mm. These hexapoles are used to refocus the beam of paramagnetic particles onto the beam axis, and thus decouple the transverse motion from the longitudinal motion that is controlled by the solenoid magnetic fields.\\
The third difference lies in the electronics. Each solenoid has a dedicated printed circuit board that can deliver two current pulses during each pass of the molecular beam. This feature allows the use of a so-called hybrid mode \cite{Cremers:PRA95:043415}, where each solenoid can be used to both accelerate and decelerate the particle beam independently. In the case where the total acceleration equals the deceleration, a packet of molecules is guided at a near-constant velocity, rejecting any molecules outside a well-defined velocity range. This guiding mode is an excellent way to produce high-density, velocity-focused beams of atoms and molecules.\\
The multistage Zeeman decelerator is constructed from five identical modules containing 20 solenoids and 19 hexapoles each. The modules are connected with an additional hexapole, ensuring that the sequence of solenoids and hexapoles is never broken. In this way, the decelerator can be built with an arbitrary length with no particle losses that would result from an interrupted solenoid-hexapole sequence. Unlike in the prototype design, the solenoids are no longer located in a 40 mm diameter tube. Instead, each module is a vacuum chamber that is machined from a solid aluminum block, and has a removable lid with an optional connection to a turbomolecular pump. In FIG. \ref{fig:DecZoom} a top-down view of the vacuum chamber is shown. For the experiments described in this paper we have built and connected five modules, for a total of 100 deceleration stages. Three out of the five chambers are equipped with commercial turbomolecular pumps (Pfeiffer Vacuum) with pumping speeds of about 250 l/s. This setup allowed us to effectively decelerate both oxygen atoms and molecules at a repetition rate of 10 Hz. The gap between the end of the decelerator and the detection point was partially bridged by an extension of the hexapole array, in order to maintain maximum particle density. To match the periodic transverse motion of the decelerated particles, the hexapole extension has the same center-to-center distance between hexapoles of 22 mm that was utilized in the decelerator.

\subsection{Detection method}
\begin{figure}[!htb]
    \centering
    \resizebox{1.0\linewidth}{!}
    {\includegraphics{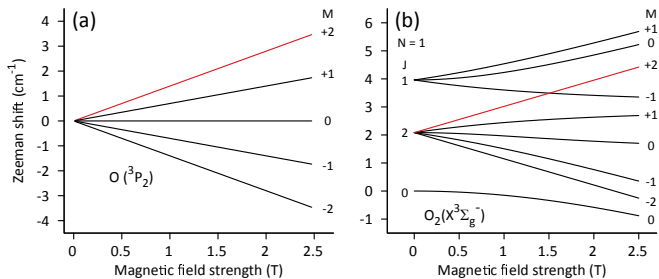}}
    \caption{The Zeeman energy level diagram of O ($^3$P$_2$) (a) and O$_2$ (X$^3\Sigma_g^-$, N = 1) (b) The low-field-seeking states with the largest Zeeman shift are indicated by the red lines (color online).}
    \label{fig:OxyEnergy}
\end{figure}

The Zeeman energy level diagrams of oxygen atoms and molecules in the ground state are shown in FIG. \ref{fig:OxyEnergy}. Both the oxygen atoms and molecules are ionized via a state-selective 2+1 REMPI transition. The oxygen atoms in the 2p$^4$ $^3$P ground state are excited to the 2p$^3$3p$^1$ $^3$P state by absorbing two photons with a wavelength of around 226 nm. A third photon of the same wavelength is used to ionize the atoms. This UV radiation was obtained from the third harmonic of an Nd:YAG pumped, pulsed dye laser containing Pyridin 1. The detection laser had an energy of approximately 0.6 mJ/pulse, a beam diameter of 5 mm and was focused using a spherical lens (f = 400 mm). The pulse duration was about 5 ns. After ionization, the positively charged ions are accelerated with electrostatic fields towards a microchannel plate (MCP) detector. Similarly, the oxygen molecules are also ionized with a 2+1 REMPI scheme, using a wavelength of around 225 nm to detect the $^3\Sigma_g^-$ ground state via a transition to the 3d Rydberg levels \cite{Yokelson:JCP97:6144}.

\section{Results}
\subsection{Oxygen atoms}
\begin{figure}[!htb]
    \centering
    \resizebox{1.0\linewidth}{!}
    {\includegraphics{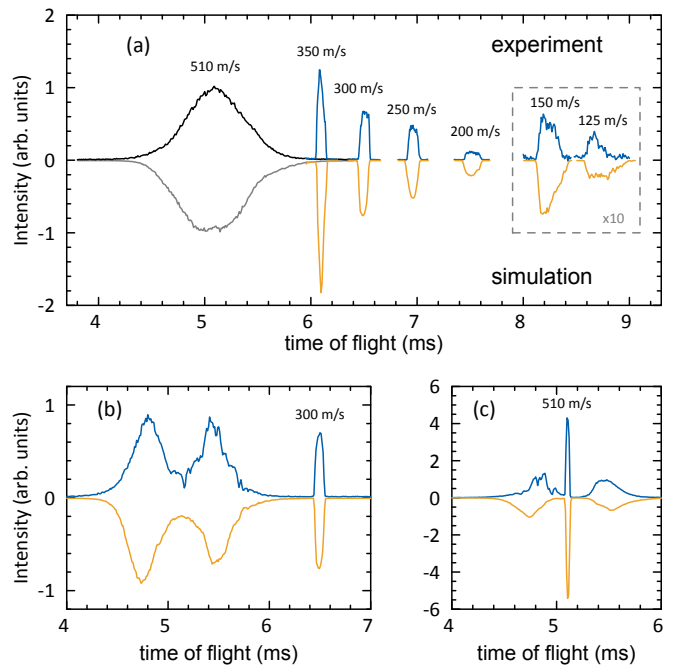}}
    \caption{(a) TOF profiles of atomic oxygen (2p$^4$ $^3$P$_2$) detected with the MCP. Experimental traces are shown in the top half of the graph, simulated profiles in the bottom half. The intensity is normalized to the non-decelerated beam (black). For the decelerated peaks, only the selected part of the beam was measured, around the final mean velocity of the deceleration sequence. The intensity of the two slowest packets has been magnified by a factor of 10, for clarity. (b) Full TOF profiles after deceleration to a final mean velocity of 300 m/s. (c) Full TOF profiles of the atomic oxygen guided at 510 m/s.}
    \label{fig:O_TOFs}
\end{figure}

In FIG. \ref{fig:O_TOFs}(a) we show time-of-flight (TOF) profiles of the beam of ground state atomic oxygen ($^3$P$_2$) using different deceleration sequences applied to the decelerator. The non-decelerated beam (black), where only transverse focusing by the permanent hexapoles affects the particle beam, is used as a reference to normalize all measurements. By operating the multistage Zeeman decelerator in deceleration mode using different values for the phase angle \cite{Cremers:PRA95:043415}, we obtain packets of O-atoms at the velocities 350 m/s, 300 m/s, 250 m/s, 200 m/s, 150 m/s and 125 m/s. At 125 m/s, the total kinetic energy of the packet has been reduced by approximately 95\%. Further deceleration will show no apparent signal peak in the TOF profile, as the packet is over-focused by the hexapoles \cite{Scharfenberg:PRA79:023410}. The full TOF profile for a beam decelerated to 300 m/s can be seen in FIG. \ref{fig:O_TOFs}(b).\\
The velocity composition of the decelerated peaks can be accurately determined with the use of particle trajectory simulations. At the higher velocities, the resulting beams are intense and already well-separated from the remaining atoms in the molecular beam. The more decelerated atoms are less focused longitudinally, resulting in reduced intensity of the peaks. Additionally, at the lowest velocities, the travel time between hexapoles can be more than the transverse oscillation time, resulting in oxygen atoms escaping the magnetic confinement of the decelerator.\\
By operating the solenoids in a double pulsed mode, the so-called hybrid mode, a range of final mean velocities close to the initial mean velocity of the beam can be produced \cite{Cremers:PRA95:043415}. This mode of operation yields the highest particle densities and allows for the transport of particles through the decelerator at constant speed. This mode is comparable to the guiding-mode used in Stark decelerators. The TOF profile that is observed when O atoms are guided at a speed of 510 m/s can be seen in FIG. \ref{fig:O_TOFs}(c). The maximum intensity is significantly larger than the decelerated beams, reflecting the greater longitudinal acceptance of this mode.

\subsection{Oxygen molecules}
\begin{figure}[!htb]
    \centering
    \resizebox{1.0\linewidth}{!}
    {\includegraphics{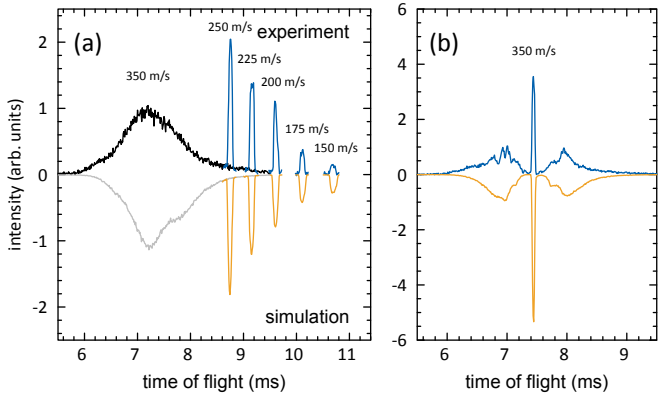}}
    \caption{(a) TOF profiles of oxygen molecules (X$^3\Sigma_g^-$) exiting the decelerator. Experimental traces are shown in the top half and simulated profiles in the bottom half. The intensity is normalized to the non-decelerated beam (black), which was produced with a mean velocity of 350 m/s. For the decelerated peaks, only the selected part of the beam is shown with corresponding mean velocity. (b) Full TOF profiles of molecular oxygen guided at 350 m/s.}
    \label{fig:O2_TOFs}
\end{figure}

The electronic ground state of O$_2$ is X$^3\Sigma_g^-$, with two unpaired spins resulting in a maximum magnetic moment for the lowest rotational state of two Bohr magneton. The magnetic moment to mass ratio is therefore only one third of that in ground state atomic oxygen. Nevertheless, with the 100-solenoid decelerator used in our experiments, this species can be effectively decelerated. However, we require an initially slower beam to reach final velocities that are well separated in time from the non-decelerated beam. Using a mixture of 10\% O$_2$ in Kr and cooling the valve to 200K, the initial mean velocity of the molecular beam is around 350 m/s. In FIG. \ref{fig:O2_TOFs}(a) we show the various decelerated peaks of oxygen molecules in the X$^3\Sigma_g^-$, J = 2 state. From the initial mean velocity of 350 m/s, we decelerate to mean velocities of 250 m/s, 225 m/s, 200 m/s, 175 m/s and 150 m/s. At 150 m/s, the kinetic energy is reduced by 80\%. In FIG. \ref{fig:O2_TOFs}(b), we show the observed TOF profile when the decelerator is used in hybrid mode to select and transport a packet with a mean velocity of 350~m/s. These results are well reproduced by simulations. We observe an over-estimation of the peak intensity in the simulations for the 350~m/s packet, and attribute this to the idealized Gaussian distributions assumed for the initial beam in the simulation, that most likely do not accurately capture the actual molecular beam distribution present in the experiment.

\subsection{State distribution}
\begin{figure}[!htb]
    \centering
    \resizebox{0.75\linewidth}{!}
    {\includegraphics{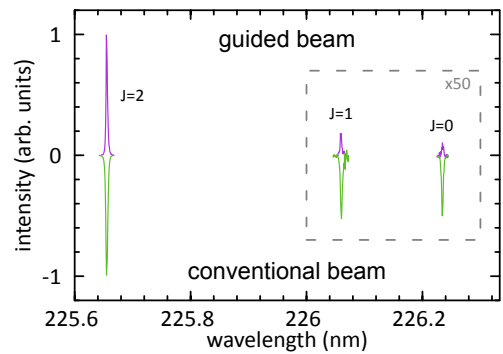}}
    \caption{The experimental REMPI transitions of atomic oxygen from the three spin-orbit levels in the electronic ground state (2p$^4$ $^3$P$_J$) to the excited state (2p$^3$3p$^1$ $^3$P$_J$). The peaks in the top half correspond to the beam that was manipulated by the Zeeman decelerator, while the spectrum in the bottom half was measured from the ELV in the secondary source chamber connected directly to the detection chamber. The intensity of the peaks for the J = 1 and J = 0 levels are magnified 50 times, for clarity. }
    \label{fig:O_SPEC}
\end{figure}

\begin{figure}[!htb]
    \centering
    \resizebox{0.75\linewidth}{!}
    {\includegraphics{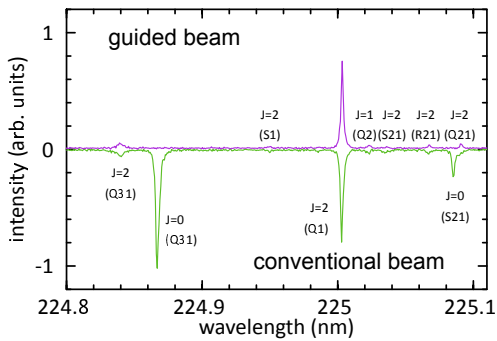}}
    \caption{The spectrum of molecular oxygen in the ground state (X$^3\Sigma_g^-$) ionized via the 3d Rydberg levels in a 2+1 REMPI transition. The top half corresponds to the beam that was manipulated by the Zeeman decelerator, while the spectrum in the bottom half was again measured from the ELV in the secondary source chamber. The signal intensity was scaled such that the peaks at 225 nm were equal in intensity. Initial J states and transitions were assigned using a simulated spectrum.}
    \label{fig:O2_SPEC}
\end{figure}

\begin{figure}[!htb]
    \centering
    \resizebox{1.0\linewidth}{!}
    {\includegraphics{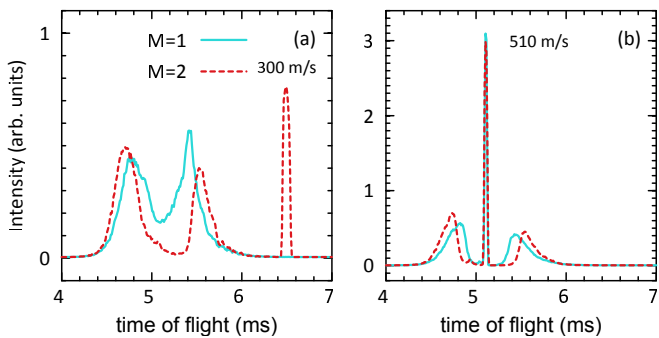}}
    \caption{Simulated TOF profiles of oxygen atoms ($^3$P$_2$) of (a) deceleration and (b) guiding sequences. The TOF profiles for the low-field-seeking states M = 1 and M = 2 are shown separately as the solid and dashed lines, respectively. Atomic oxygen in the M = 0 state is not shown, because it is not focused by the hexapole array and therefore negligible in intensity. High-field-seeking M = -1, -2 states are defocused by the hexapole array, and therefore do not reach the end of the decelerator.}
    \label{fig:O_states}
\end{figure}

For both atomic and molecular oxygen, the Zeeman decelerator can be used to select a single quantum state, typically the state with the largest magnetic dipole moment. In atomic oxygen this is the $^3$P$_2$ spin-orbit component of the $^3$P$_J$ electronic ground state and for molecular oxygen this is the J = 2 rotational level of the X$^3\Sigma_g^-$ state in the N = 1 rotational ground state. Note that following common convention in nomenclature, $J$ is used here to indicate the spin-orbit component of the $^3$P ground state of O, as well as to indicate the spin-rotational energy levels of O$_2$. \\
In FIG. \ref{fig:O_SPEC} we show the measured REMPI transitions of oxygen atoms from the electronic ground state. In the top half, the spectrum is measured after the beam has been guided by the multistage Zeeman decelerator at a speed of 510~m/s. The spectrum in the bottom half was measured without the decelerator. For this measurement, the ELV was placed in a second source chamber connected to the detection chamber, with a 3 mm diameter skimmer (Beam Dynamics) to collimate the beam. In both beams, the atoms mostly occupy the ground state (J=2), with less than 1\% of the signal intensity corresponding to population in the J = 1 and J = 0 spin-orbit states. In the guided beam, the relative population of the J=2 state is enhanced due to the state-selecting properties of the multistage Zeeman decelerator. In FIG. \ref{fig:O2_SPEC} the spectrum of O$_2$ is shown, again obtained from the supersonic beam with and without the decelerator to guide the beam. In this case, the Zeeman decelerator is operated to select and guide a packet with a mean velocity of 350~m/s. Since the J = 0 rotational energy level is the absolute ground state, it is the most populated state in the conventional beam. However, after passing through the multistage Zeeman decelerator, this state is absent from the spectrum. This is due to the slight high-field seeking Zeeman shift of this state (as seen in FIG. \ref{fig:OxyEnergy}), resulting in transverse defocusing by the hexapole array.\\
With the decelerator we select the most low-field-seeking M = 2 projection of the J = 2 state, while also transmitting the low-field-seeking M = 1 state. However, due to an insufficient magnetic field between the end of the decelerator and the detection area, the M-states become degenerate and redistribute before they are detected. Experimentally, they become indistinguishable. However, simulations allow us to identify contributions of different M states to the TOF profiles. This can be used to predict the state-purity of the decelerated beams. As an example, TOF profiles of atomic oxygen in the M = 1 and M = 2 states are shown separately in FIG. \ref{fig:O_states}, using the simulated deceleration and guiding profiles from FIG. \ref{fig:O_TOFs}. After deceleration to 300 m/s, oxygen atoms in the M = 2 state are shown to be completely separated from those in the M = 1 state. We can also conclude that oxygen atoms in the J = 1 state are filtered out of this decelerated beam, since their spin-orbit level does not have a M = 2 projection. With the guiding sequence, both low-field-seeking M-states are transmitted, though the accepted velocity spread is slightly larger for oxygen atoms in the M = 2 state.

\section{Conclusion}

We presented a new 3-meter long multistage Zeeman decelerator consisting of 100 solenoids and 100 magnetic hexapoles, based on a short prototype design presented recently. The decelerator features a unique solenoid design that allows for effective cooling and relatively cheap, low-voltage electronics. The decelerator is fully modular without interrupting the hexapole-solenoid sequence, has excellent vacuum properties and can be robustly operated at 10 Hz. Beams of oxygen atoms are decelerated for the first time, producing packets of O atoms in a single spin-orbit state and with a well-defined and tunable velocity. We further characterized the performance of the decelerator using a seeded beam of molecular oxygen. The results presented here show that the Zeeman decelerator concept of using an alternating array of hexapoles and solenoids is truly phase stable, and decelerators using this concept may in principle be constructed with arbitrary length.\\
The decelerator presented here allows for the production of high-density, state-selected and velocity-tunable packets of atoms or molecules. The variety of species that can be used range from the relatively light species O, O$_2$ and NH, to heavier species like CaF, SrF and MnH. These species are prime candidates for controlled and low-temperature scattering studies. For instance, the velocity-controlled packets of O atoms demonstrated here are ideally suited to study low-energy inelastic or reactive collisions for a variety of benchmark systems. Theoretical predictions for inelastic O + He \cite{Lique:MNRAS474:2313} and O + H \cite{Abrahamsson:AJ654:1171} that either excite or de-excite O atoms from one spin-orbit state to another suggest that rich quantum phenomena like scattering resonances can be explored at low energies and at energies around the threshold for spin-orbit excitation \cite{Ma:CPL213:269}. The O ($^3$P)+ OH(X$\,^2\Pi$) $\rightarrow$ O$_2$(X$\,^3\Sigma_{g}^{+}$) + H ($^2$S) reaction is the inverse of "the single most important combustion reaction" \cite{Xie:JCP126:074315}, and is of pivotal importance to gas phase processes in atmospheres and in interstellar clouds. Theoretically, a spectacular increase in reaction rate for decreasing collision energies was predicted \cite{Lin:JCP128:014303}, although major discrepancies between theory and experiment exist at high collision energies for which experimental data is available \cite{Ma:JCP133:054302}. The controlled packets of Zeeman-decelerated O atoms that are now available offer exciting prospects to experimentally study these processes with unprecedented sensitivity and precision.

\section{Acknowledgments}
The research leading to these results has received funding from the European Research Council under the European Union's Seventh Framework Programme (FP7/2007-2013/ERC grant agreement nr. 335646 MOLBIL). This work is part of the research program of the Netherlands Organization for Scientific Research (NWO). We thank Peter Claus for the solenoid concept and Gerben Wulterkens for the final design of the multistage Zeeman decelerator.


\end{document}